\documentclass[%
 preprint,
 amsmath,amssymb,
 aps,
]{revtex4-2}
\usepackage{makecell}
\usepackage{graphicx}
\usepackage{dcolumn}
\usepackage{bm}
\usepackage{hyperref}

\usepackage{float}
\begin{document}

\preprint{APS/123-QED}

\title{Excitonic Condensate in Flat Valence and Conduction Bands of Opposite Chirality}

\author{Gurjyot Sethi\textsuperscript{1}}
\author{Martin Cuma\textsuperscript{2}}%
\author{Feng Liu\textsuperscript{1}}
\affiliation{%
 \textsuperscript{1}Department of Materials Science and Engineering, University of Utah, Salt Lake City, Utah 84112, USA\\
\textsuperscript{2}Center for High Performance Computing, University of Utah, Salt Lake City, Utah 84112, USA\\
}%

\date{\today}

\begin{abstract}
Excitonic Bose-Einstein condensation (EBEC) has drawn increasing attention recently with the emergence of 2D materials. A general criterion for EBEC, as expected in an excitonic insulator (EI) state, is to have negative exciton formation energies in a semiconductor. Here, using exact diagonalization of multi-exciton Hamiltonian modelled in a diatomic Kagome lattice, we demonstrate that the negative exciton formation energies are only a \textit{prerequisite} but \textit{insufficient} condition for realizing an EI. By a comparative study between the cases of both a conduction and valence flat bands (FBs) versus that of a parabolic conduction band, we further show that the presence and increased FB contribution to exciton formation provide an attractive avenue to stabilize the EBEC, as confirmed by calculations and analyses of multi-exciton energies, wave functions and reduced density matrices. Our results warrant a similar many-exciton analysis for other known/new candidates of EIs, and demonstrate the FBs of opposite parity as a unique platform for studying exciton physics, paving the way to material realization of spinor BEC and spin-superfluidity.
\end{abstract}

\maketitle


Excitonic Bose-Einstein condensate (EBEC), first proposed in 1960s \cite{1,2,3,4}, has drawn recently increasing interest with the emergence of low-dimensional materials where electron screening is reduced leading to increased exciton binding energy ($E_b$) \cite{5,6}. In 1967, Jerome, et. al. \cite{7}, theoretically presented the possibility of an excitonic insulator (EI) phase in a semi-metal or a narrow gap semiconductor \cite{7,8,9,10}. It was shown that the hybridization gap equation for excitonic condensate order parameter has non-trivial solutions, when $E_b$ exceeds the semiconductor/semi-metal band gap ($E_g$). In deep semi-metallic regime, this gap equation can be solved in analogy to Bardeen-Cooper-Schiffer (BCS) superconductor theory \cite{7,11}. Due to strong screening of Coulomb potential by the carriers in a semi-metal, there exists an electron-hole plasma which forms a condensate of weakly paired electrons and holes at low temperature. On the other hand, in a semiconductor regime, preformed excitons may condense to form a BEC at low temperatures \cite{7,11}.\par
This has led to significant theoretical \cite{6,12,13,14,15,16,17,18,19} and experimental \cite{20,21,22,23,23ky,24,25,26,27,28,29,30,31} investigations into finding an EI state in real materials. Especially, the EI state in a semiconductor provides an alternative route to realizing EBEC instead of targeting materials with long-lifetime excitons, such as optically inactive excitons in bulk Cu\textsubscript{2}O \cite{32,33,34,35,36} and indirect excitons in coupled quantum wells \cite{5,37,38}. It is worth mentioning that excitonic condensation has been reported in double layer 2D heterostructures \cite{39,40,41,42,43,44,45,46,47,48,49}, where electrons and holes are separated into two layers with a tunneling barrier in between, and double-layer quantum Hall systems \cite{50,51,52,53,54} have been shown to exhibit excitonic condensation at low temperature under a strong magnetic field. On the contrary, EIs are intrinsic, i.e., excitonic condensate stabilizes spontaneously at low temperature without external fields or perturbations.\par
However, experimental confirmation of EI state remains controversial \cite{20,21,22,23,23ky,24,25,26,27,28,29,30,31}, mainly because candidate EI materials are very limited. On the other hand, some potential candidate EIs have been proposed by state-of-the-art computational studies \cite{6,12,13,14,15,16,17,18,19}, based on calculation of single exciton formation energy. It is generally perceived that if single exciton $E_b$ exceeds the semiconductor $E_g$, the material could be an EI candidate. But the original mean-field two-band model studied in Ref. \cite{7} includes inter/intra band interactions, leading to a non-trivial condensation order parameter, which indicates the importance of multi-exciton interactions. Hence, in order to ultimately confirm new EI candidates, it is utmost necessary to analyze and establish the stabilization of multi-exciton condensate with quantum coherency in the parameter space of multiple bands with inter/intra band interactions, beyond just negative formation energy for single or multiple excitons.\par
In this Letter, we perform multi-exciton wave function analyses beyond energetics to directly assess EBEC for a truly EI state, namely a macroscopic number of excitons (bosons) condensing into the same single bosonic ground state \cite{55,56,57,58}. Especially, we investigate possible stabilization of EBEC in a unique type of band structure consisting of a pair of valence and conduction flat bands (FBs) of opposite chirality. These so-called yin-yang FBs were first introduced in a diatomic Kagome lattice \cite{59,60} and have been studied in the context of metal-organic frameworks \cite{61} and twisted bilayer graphene \cite{62}. Recently, it was shown that such FBs, as modelled in a superatomic graphene lattice, can potentially stabilize a triplet EI state due to reduced screening of Coulomb interaction \cite{6}. However, similar to other previous computational studies \cite{16,17,18,19}, the work was limited to illustrating the spontaneity of only a single exciton formation with a negative formation energy. Here, using exact diagonalization (ED) of a many-exciton Hamiltonian based on the yin-yang FBs, in comparison with the case of a parabolic conduction band, we demonstrate that “$E_b>E_g$” is actually only a necessary but \textit{insufficient} condition for realizing an EI state. While both systems show negative multi-exciton energies, only the former was confirmed with quantum coherency from the calculation of off-diagonal long-rang order (ODLRO) of the many-exciton Hamiltonian. Furthermore, we show that with the increasing FBs contribution to exciton formation, the excitons, usually viewed as composite bosons made of electron-hole pairs, can condense like point bosons, as evidenced from the calculated perfect overlaps between the numerical ED solutions with the analytical form of ideal EBEC wave functions.\par
\begin{figure}[h]%
\centering
\includegraphics[width=0.7\textwidth]{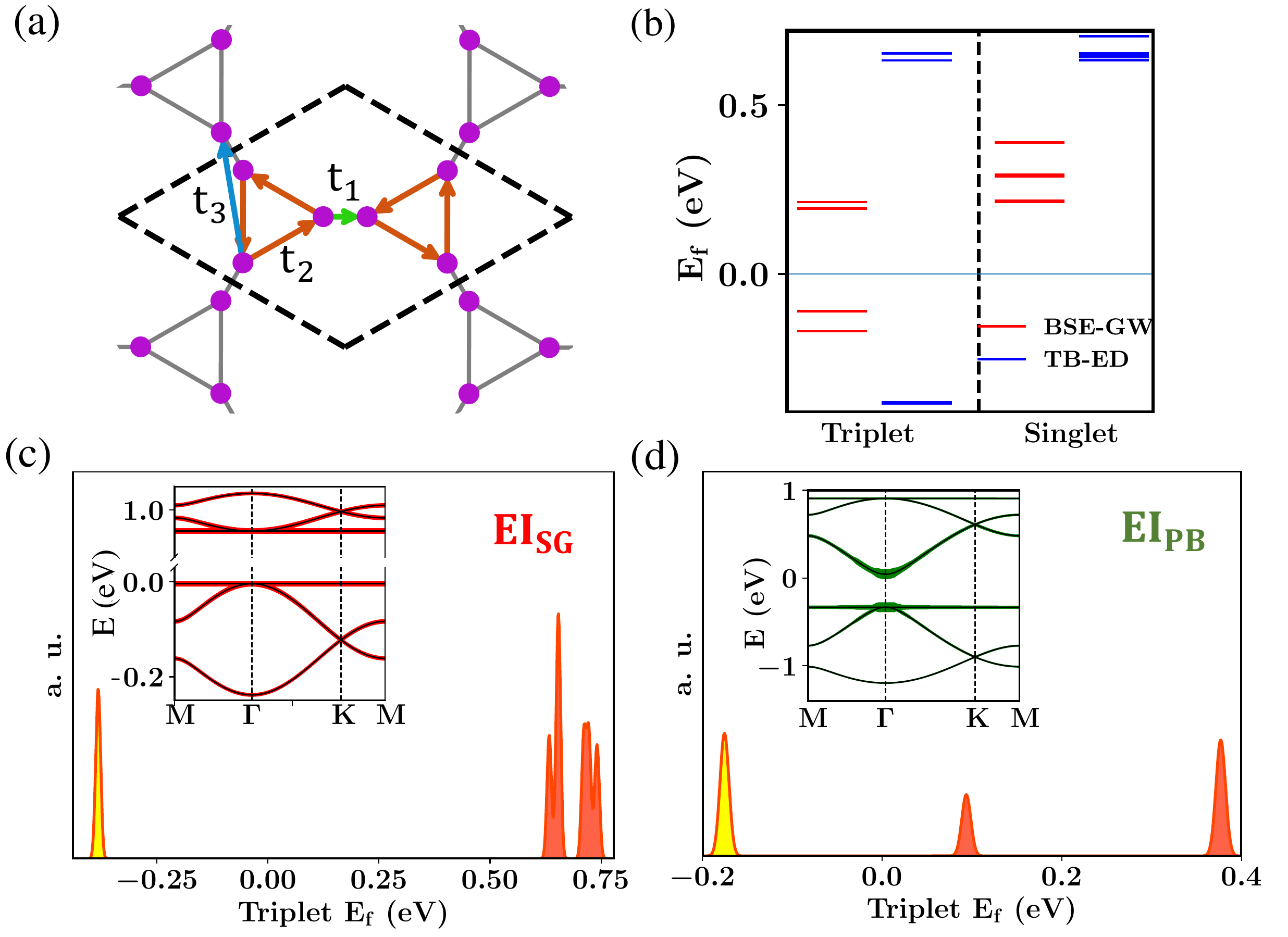}
\caption{(a) Schematic of diatomic Kagome lattice with first three NN hopping integrals labelled as $t_1$, $t_2$, and $t_3$, respectively. (b) Single exciton $E_f$ calculated using ED (blue bars) compared with GW-BSE results \cite{6} (red bars) for $EI_{SG}$. (c), and (d) Triplet excitonic density of states for $EI_{SG}$, and $EI_{PB}$ respectively. Excitonic states with negative and positive formation energies are shown in yellow and orange, respectively. Inset shows the band excitation contributions to the first triplet level, indicated by the width of bands in red for $EI_{SG}$ ((c)) and green for $EI_{PB}$ ((d)), respectively.}\label{fig1} 
\end{figure}

A tight-binding model based on diatomic Kagome lattice is considered for the kinetic energy part of the Hamiltonian, as shown in Fig. ~\ref{fig1}(a). Our focus will be on comparing the many-excitonic ground states of superatomic graphene lattice (labelled as $EI_{SG}$), which is already known to have a negative single exciton formation energy \cite{6}, and the ground states of a model system (labelled as $EI_{PB}$) with a parabolic conduction band edge, in order to reveal the role of FBs in promoting an EI state. The interatomic hopping parameters for the two systems are: $t_1=0.532$ eV; $t_2=0.0258$ eV; $t_3=0.0261 eV$ for $EI_{SG}$, benchmarked with density-functional-theory (DFT) results \cite{6,63}, and $t_1=0.62$ eV; $t_2=0.288$ eV; $t_3=0.0$ eV for $EI_{PB}$. An interesting point to note here is that for $EI_{SG}$, $t_2<t_3$. This is an essential condition to realize yin-yang FBs in a single-orbital tight-binding model as has been discussed before, which can be satisfied in several materials \cite{59,60,61}. The insets in Fig. ~\ref{fig1}(c) and ~\ref{fig1}(d) show the band structures for $EI_{SG}$ and $EI_{PB}$, respectively. Coulomb repulsion between electrons is treated using an extended Hubbard model as
\begin{align}
H=H_{kin}+H_{int}&=\sum_n\sum_{<r,r'>_n}t_nc_r^{\dagger}c_{r'}+\nonumber\\
&+\sum_n\sum_{<r,r'>_n}V_nc_r^{\dagger}c_rc_{r'}^{\dagger}c_{r' },\label{eqn1}
\end{align}
where $t_n$ is the $n^{th}$ nearest-neighbor (NN) hopping parameter, and $V_n$ is $n^{th}$  NN Hubbard parameter. Each of the $V_n$  is calculated using the Coulomb potential, $U(r>r_o)=e^2/(4\pi\epsilon\epsilon_o r)$, with a very low dielectric constant ($\epsilon\sim1.02$) due to the presence of FBs in a 2D lattice \cite{6} and a cutoff ($r_o$) for onsite interactions. The Hubbard interaction terms are projected onto all three conduction and valence bands. Spin indices in the Hamiltonian are omitted. We distinguish triplet and singlet excitonic states by the absence and presence of excitonic exchange interaction, respectively \cite{63,64}. The Hamiltonian is exactly diagonalized for a finite system size ($2\times3$) for converged results \cite{63}, which includes 36 lattice sites (equivalent to a $6\times6$ trigonal lattice) with 18 electrons for a half-filled intrinsic semiconductor. With $N_{eh}$ number of electrons (holes) in conduction (valence) bands, exciton population (EP) is defined as $N_{eh}$ divided by the total number of allowed reciprocal lattice points (i.e., $2\times3 = 6$). Throughout this work we focus on the ground state of Eqn. ~\ref{eqn1} with varying EPs.\par
We first calculate the energies and wavefunctions for a single exciton, i.e., $N_{eh}=1$ (EP=1/6), to benchmark the single-exciton results of $EI_{SG}$ with those obtained using first-principles GW-BSE method for this lattice \cite{6}. Importantly, our model calculation results, especially the trends of exciton levels, match very well with GW-BSE (Fig. ~\ref{fig1}(b), Fig. S2 \cite{63}). One clearly sees in Fig. ~\ref{fig1}(b) for $EI_{SG}$ that the formation of triplet exciton is spontaneous with a negative formation energy ($E_f$), while that of singlet is positive. These key agreements validate our model for further analysis. In Fig. ~\ref{fig1}(c) and ~\ref{fig1}(d), we plot triplet excitonic density of states for $EI_{SG}$ and $EI_{PB}$, respectively. Both systems have a negative lowest triplet $E_f$, indicative of the possibility that both systems can be a triplet EI. The insets of Fig. ~\ref{fig1}(c) and ~\ref{fig1}(d) show the band excitation contribution to the lowest triplet exciton level. For $EI_{SG}$ (Fig. ~\ref{fig1}(c)), as has been shown before by GW-BSE method \cite{6}, all three band excitations contribute almost equally throughout the entire Brillouin zone (BZ). In contrast, for $EI_{PB}$ (Fig. ~\ref{fig1}(d)), the $\Gamma$-point excitation contributes the most due to the presence of parabolic conduction band edge with band minimum at $\Gamma$. In this study, we will focus on triplet excitons, which have negative $E_f$ in both systems, so unless otherwise specified, excitons below mean triplet excitons.\par
\begin{figure}[h]%
\centering
\includegraphics[width=0.7\textwidth]{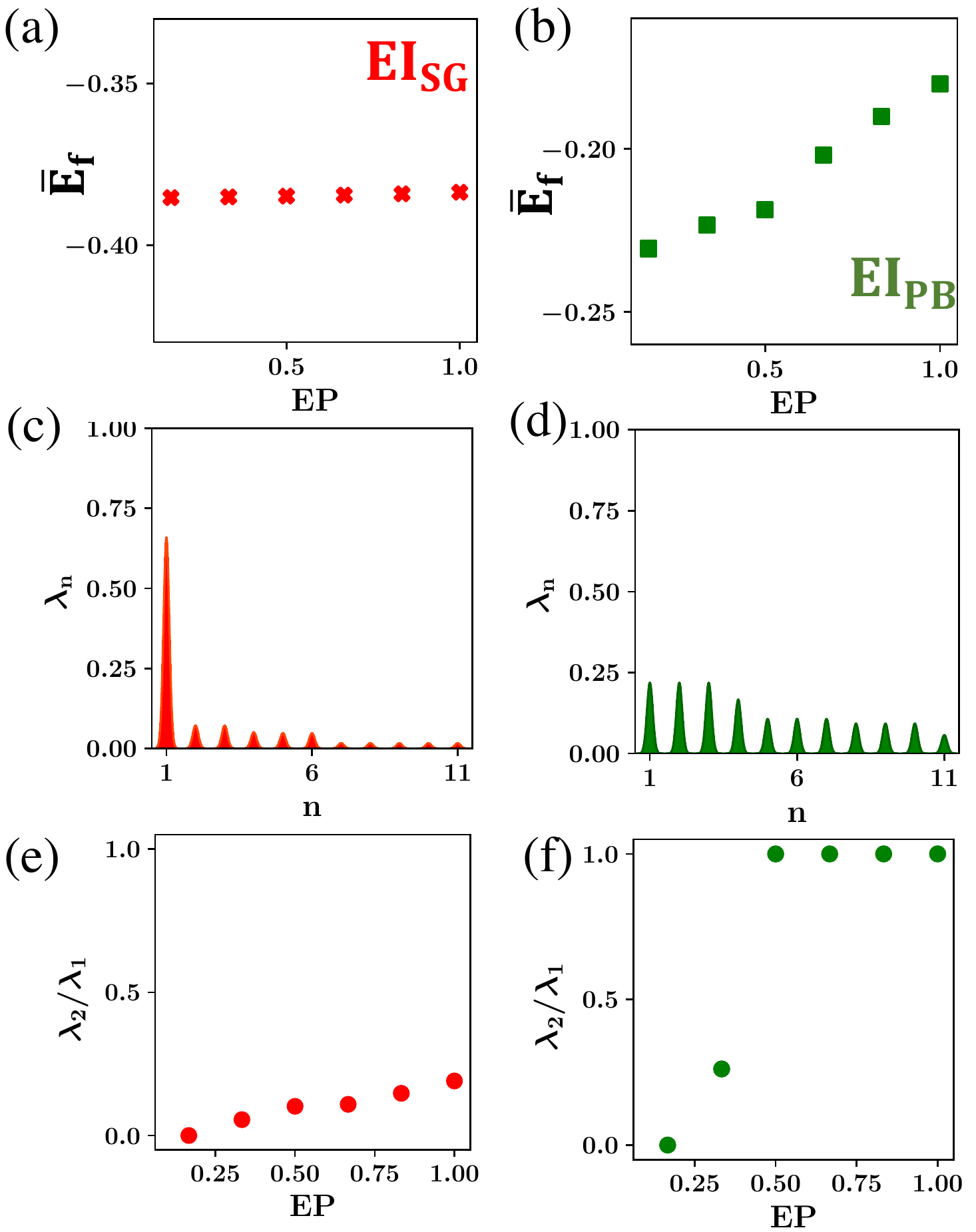}
\caption{(a) $\overline{E}_f$ of the ground-state multi-triplet-exciton states at multiple EPs for $EI_{SG}$. (b) Same as (a) for $EI_{PB}$. Scale of plots in (a) and (b) is kept identical for comparison. (c) First few largest normalized eigenvalues ($\lambda_n$) of reduced two-body density matrix calculated for the ground-state multi-triplet-exciton wave functions of $EI_{SG}$ at EP $\sim$ 0.67. (d) Same as (c) for $EI_{PB}$. (e) Ratio $\lambda_2/\lambda_1$ plotted at various EPs as an indicator of fragmentation in the ground states of $EI_{SG}$. (f) Same as (e) for $EI_{PB}$.}\label{fig2}
\end{figure}

Next, we discuss many-exciton calculations. A BEC superfluid flows with minimal dissipation \cite{56}. Statistically, the BEC state is characterized with a Poisson particle distribution manifesting a non-interactive nature \cite{65}. In other words, even in the presence of interactions, there should be a minimal change in the average formation energy ($\overline{E}_f$) of a superfluid when more particles are condensed. To reveal such effect of exciton-exciton interactions on spontaneity of exciton formation and condensation, we exactly diagonalize (~\ref{eqn1}) for $N_{eh}>1$. In Fig. ~\ref{fig2}(a), and Fig. ~\ref{fig2}(b), we show the average ground-state $\overline{E}_f$ of excitons with increasing EP for $EI_{SG}$, and $EI_{PB}$, respectively, namely the multi-exciton ground-state $E_f$ divided by $N_{eh}$. Note that both plots have the same scale to facilitate a direct comparison.\par
In both cases, the ground-state excitons have negative formation energies at all EPs, but importantly the nature of exciton-exciton interactions are different. For $EI_{SG}$, the excitons experience a very slight repulsive exciton-exciton interaction, indicated by a very small positive slope of their $\overline{E}_f$ curve (Fig. ~\ref{fig2}(a)). From EP = 0.17 to EP = 1.0, $\overline{E}_f$ increases by only 0.47\%. Differently for $EI_{PB}$, excitons experience a strong repulsion from each other (Fig. ~\ref{fig2}(b)); $\overline{E}_f$ increases by 21.9\% from EP = 0.17 to EP= 1.0. Consequently, we make the following inferences. First, the excitons in $EI_{SG}$ are likely forming a BEC superfluid in the ground state because the effect of exciton-exciton interactions on $\overline{E}_f$ is negligible. In the sense of weak exciton-exciton repulsion, the low-lying excitons for $EI_{SG}$ appear like composite bosons, similar to weakly repulsive bosons in helium-II \cite{66}. Secondly, the existence of negative exciton formation energy alone is possibly insufficient to establish a coherent BEC state. The multi-excitonic ground state of $EI_{PB}$ has also negative formation energies, but judging from the strong exciton-exciton interaction excitons seem to unlikely form a condensate. In order to confirm this argument, however, one has to further assess directly the nature of exciton-exciton interaction and confirm quantum coherence of multi-exciton wavefunctions as we do next.\par
Since excitons are composite bosons made of electron-hole pairs like Cooper pairs of two electrons, we calculate eigenvalues of reduced two-body density matrix as a definitive signature of EBEC based on the concept of off-diagonal long-range order (ODLRO), which was first introduced to characterize superfluidity of Cooper pairs \cite{66,67}. Similarly, the reduced two-body density matrix for excitons can be written as \cite{63},
\begin{equation}
\rho^{(2)}(k,k';\overline{k},\overline{k}')=<\Psi\vert \psi_c^{\dagger}(k)\psi_v(k')\psi_v^{\dagger}(\overline{k}')\psi_c(\overline{k})\vert\Psi>,\label{eqn2}
\end{equation}
where $\psi_{c(v)}^{\dagger}(k)$ creates a conduction (valence) electron at reciprocal lattice point $k$, and $\vert\Psi>$ is the many-exciton wavefunction.We calculate the eigenvalues of $\rho^{(2)}$ and normalize it by $N_{eh}$ as a function of EP, then the existence of a single normalized eigenvalue close to 1 is a signature of EBEC \cite{63}. We also calculate the ratio of the first two eigenvalues to check for fragmentation \cite{68} of multi-exciton ground state. Ideally, this ratio should be close to zero; if it is close to 1, it indicates fragmentation of the condensate.\par
In Fig. ~\ref{fig2}(c), we plot the eigenvalue spectra ($\lambda_n$) of $\rho^{(2)}$ for the many-body ground state of excitons for $EI_{SG}$ at EP $\sim$ 0.67, in a descending order, i.e., $\lambda_n$ being the $n^{th}$ largest eigenvalue. Similar results are found for all EPs (see Fig. S4 \cite{63}). Clearly, there appears a high degree of condensation for EP $\sim$ 0.67. It can also be seen from Fig. ~\ref{fig2}(e), where the ratio $\lambda_2/\lambda_1$, indicative of fragmentation of the condensate, is very low for all EPs. For comparison, in Fig. ~\ref{fig2}(d), we plot the $\lambda_n$ spectra for the many-body ground state of excitons for $EI_{PB}$ at EP $\sim$ 0.67. Again, similar results are found for other EPs (see Fig. S5 \cite{63}). The excitons in this case, however, are clearly not condensing even though they have also negative $E_f$ as shown in Fig. ~\ref{fig1}(d) and ~\ref{fig2}(b). It can be seen from Fig. ~\ref{fig2}(f) that the multi-exciton ground state is completely fragmented as $\lambda_2/\lambda_1$ goes to 1 with the increasing EP. Therefore, by examining the nature of multi-exciton wave functions we conclude that the condition of “$E_b>E_g$”, as satisfied in both cases, is only a necessary but insufficient condition for EI state. Also, it indicates that the superatomic graphene can be a promising real candidate material for realizing a true EI with excitonic coherence for all EPs.\par
\begin{figure}%
\centering
\includegraphics[width=0.7\textwidth]{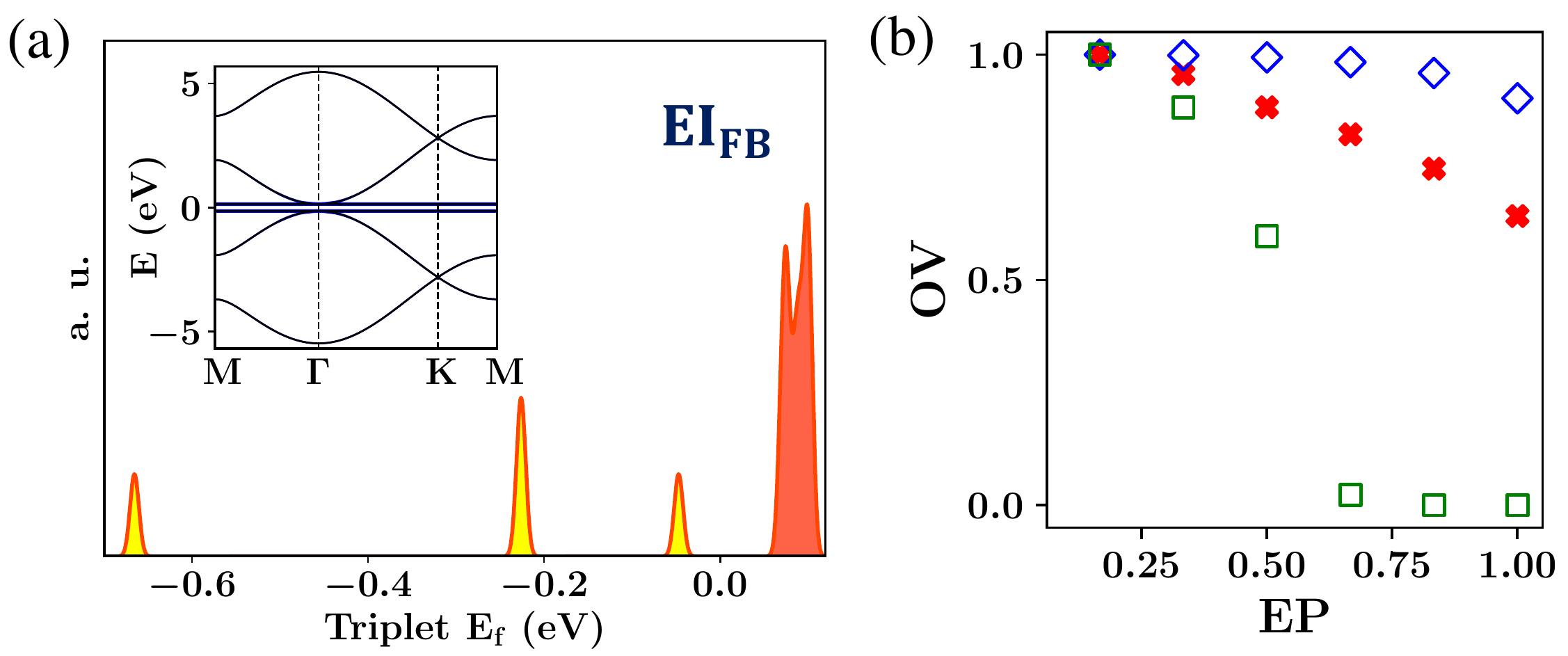}
\caption{(a) Same as Fig. ~\ref{fig1}(c) and ~\ref{fig1}(d) for $EI_{FB}$. (b) Overlaps of ED calculated wave function with the BEC wave function of the form given by Eqn. (~\ref{eqn3}) for the ground states of $EI_{SG}$ (red crosses), $EI_{FB}$ (blue diamonds), and $EI_{PB}$ (green squares), at various EPs.}\label{fig3}
\end{figure}
Moreover, the above comparative study suggests that FB is preferable to enhance exciton coherence, as opposed to parabolic band. Interestingly, in our tight-binding model of a diatomic Kagome lattice, it is possible to increase the relative FB contribution to exciton formation by tuning the hopping parameters. Specifically, we can reduce the band gap between the yin and yang FB \cite{63} to increase the contribution of FB excitations to the lowest excitonic state, as exemplified in Fig. ~\ref{fig3}(a) using the hopping parameters: $t_1=1.92$ eV; $t_2=0.0$ eV; $t_3=0.93$ eV (labelled as $EI_{FB}$), where we plot the single excitonic energy levels and band excitation contributions (inset) to the lowest triplet level of $EI_{FB}$. Note that even with a small $E_g$ in this case, excitons have a large $E_b$ because FBs host massive carriers, leading to a very small dipole matrix element between them \cite{6}, which enables a low-band-gap system to still have a very low screening \cite{69}. The lowest exciton level of $EI_{FB}$ has a negative $E_f$ and FB excitations contribute the most to this level.\par
Similar to the above analyses for $EI_{SG}$ and $EI_{PB}$, we have used ODLRO calculation to verify that multi-exciton ground state of $EI_{FB}$ is an EI state \cite{63} with a slight fragmentation at higher EP (see Fig. S6, S7 \cite{63}). On the other hand, due to large FB contributions, the exciton-exciton interaction will be affected, such as by increasing the electron-hole overlap through the compact localized states of yin-yang FBs in real space to make excitons more “compact”. Then excitons in the case of $EI_{FB}$ may behave more like a point boson and condense in a form to also resemble a one-body BEC state. Next we illustrate this possibility by further analyzing the ground-state many-body wave functions obtained from ED calculations. One can use the single exciton wave function, calculated with $N_{eh}=1$, to form an ideal N-exciton BEC wavefunction in the form of \cite{55,56,63,65},
\begin{equation}
\vert\phi_{BEC}>=\frac{1}{\Omega}[b_{exc}^{\dagger}]^N\vert0>,\label{eqn3}
\end{equation}
where $b_{exc}^{\dagger}$ is the creation operator for the single triplet level obtained from ED with $N_{eh}=1$, $\Omega$ is the normalization constant and $N$ is the number of electrons (holes) in conduction (valence) bands. Let $\vert\phi_{ED}>$ be the ED solution with $N$ electrons (holes) in conduction (valence) bands. In Fig.~\ref{fig3}(b) we show the overlap, $OV=\vert<\phi_{BEC}\vert\phi_{ED}>\vert$ for the multi-exciton ground state of $EI_{FB}$, $EI_{SG}$ and $EI_{PB}$ with increasing EP.\par
The BEC-ED overlaps are very close to one for the ground state of $EI_{FB}$ at all EPs (blue diamonds in Fig. ~\ref{fig3}(b)), indicating that excitons in this case are condensing into a one-body BEC form. In contrast, for the case of $EI_{SG}$ (red crosses in Fig. ~\ref{fig3}(b)), the overlap monotonically decreases with the increasing EP, implying that the ground state is not a BEC state of the form in Eqn. (~\ref{eqn3}), especially at higher EP. On the other hand, we already showed above from the ODLRO (Fig. ~\ref{fig2}) that the excitons of $EI_{SG}$ do form a condensate, albeit in a different form. In this sense, the ground-state excitons of $EI_{SG}$ behave and condense as composite bosons of electron-hole pairs like Cooper pairs; while those of $EI_{FB}$ behave and condense as one-body bosons as if without internal structure more like cold atoms. For the case of $EI_{PB}$ (green squares in Fig. ~\ref{fig3}(b)), the overlap stays much less than unity at all EPs, so that excitons are not condensing in either one-body or composite form (Fig. ~\ref{fig2}(f)), which is not surprising given the strong exciton-exciton interaction (Fig. ~\ref{fig2}(b)).\par
We point out that the presence and large contribution of FB excitations to the excitonic level appear to be preferable for EBEC. This is clearly reflected by comparing the three cases studied. In the case of $EI_{PB}$ with a parabolic conduction band edge, the lowest triplet level is largely contributed by only $\Gamma$-point excitation (Fig. ~\ref{fig1}(d)). Excitons fail to form a BEC at all EPs (Fig. ~\ref{fig2}(d), ~\ref{fig2}(f) and Fig. S5 \cite{63}) despite having negative formation energies. In the case of $EI_{SG}$ with both a flat valence and conduction band edge, the lower level is contributed by FBs at all k-points along with other parabolic bands (Fig. ~\ref{fig1}(c)). Excitons condense into a composite form at all EPs (Fig. ~\ref{fig2}(c), ~\ref{fig2}(e) and Fig. S4 \cite{63}), but lose the coherence in the simple ideal form of Eqn. ~\ref{eqn3} as EP increases (Fig. ~\ref{fig3}(b)). In the case of $EI_{FB}$ with further increase of FB excitations to the ground-state exciton level (Fig. ~\ref{fig3}(a)), exciton condense into the ideal form like point bosons (Fig. ~\ref{fig3}(b)). In general, the presence of FB appears to help in improving exciton coherency.  This is consistent with a recent study \cite{70} showing the stability of condensate hosted by FB even in the limit of no interactions in stark contrast to parabolic band.\par
Last but not least, FBs are considered as the solid-state analogue of Landau levels (LLs) in free electron gas under strong magnetic fields \cite{71,72,73}, which led to the realization fractional Chern insulators \cite{74,75,76}. The lattice model studied in this work with yin-yang FBs can similarly be viewed as a kind of solid-state analogue of quantum Hall bilayer (QHB), with the conduction and valence FBs representing individual LL in each layer. In a QHB excitonic condensation occurs at the total filling $v_T=1$ \cite{50} when the ground-state wave function resembles the Halperin’s (1,1,1) state \cite{77}. Although there have been recent efforts of stabilizing this state in a single topological FB partially filled with spin-up and spin-down fermions \cite{78}, yin-yang FBs should provide a more natural way of realizing anomalous QHB states without magnetic field. Our work here instigates further investigation into this analogy. One intriguing point to note is that QHB systems have access to only singlet excitons since the magnetic field is in the same direction for both layers, while yin-yang FBs of opposite parity allow also for triplet excitons, which can lead to realization of exotic new phases like fractional excited spin Hall effect. In addition, the stabilization of triplet EI state, as illustrated here for diatomic Kagome lattice, paves the way towards material realization of exotic phases like spin-1 bosonic condensate \cite{79,80} and spin superfluidity \cite{81,82}. In summary, our work has significantly enriched the FB and excitonic physics, and demonstrated convincingly the potential of FB-materials for realizing EBEC while providing a computational framework to perform multi-exciton analysis for quantum coherency in other known/new candidates of EIs.\par
This work is supported by US Department of Energy-Basic Energy Sciences (Grant No. DE-FG02- 04ER46148). All calculations were done on the CHPC at the University of Utah.

\section*{Supplementary Material}
\section{Methods}

\subsection*{Many-exciton calculation setup}
As shown by Eqn. 1 in the main text, kinetic energy part of the many-exciton Hamiltonian is based on a tight-binding (TB) model of diatomic Kagome lattice. We note in Eqn. 1 that we have adopted an extended Hubbard model for the Coulomb interaction between electrons, instead of using a distance-dependent potential, because in the lattices we are considering here the electron-hole distances are much smaller than the lattice constant \cite{6} so that using the latter approach would have made these calculations highly expensive due to the requirement of complete two-point Fourier transform \cite{64}, unlike in materials like MoS\textsubscript{2} \cite{83,84} where a one-point Fourier transform is sufficient. Also, we use a small dielectric constant to model screening for multiple band gaps since in yin-yang Kagome band structure, there is highly reduced screening due to the presence of flat conduction and valence band \cite{6}. As long as the flat bands (FBs) are present we can use a very low dielectric constant for multiple band gaps.\par
We use a bands-projected interaction given by
\begin{align}
H_{int}^{proj}&=\frac{V_n}{N}\sum_{k_i}\delta^{2\pi}_{k_1+k_3-k_2-k_4}\sum_{<x,y>_n}V^{xy}_{k_1,k_2,k_3,k_4} \nonumber\\
&\qquad\sum_{\alpha_i}u^*_{xk_1,\alpha_1}u_{xk_2,\alpha_2}u^*_{yk_3,\alpha_3}u_{yk_4,\alpha_4}c_{\alpha_1k_1}^{\dagger}c_{\alpha_2k_2}c_{\alpha_3k_3}^{\dagger}c_{\alpha_4k_4},\label{eqn2}
\end{align}
where $V^{xy}$ is phase factor acquired by the pair $<x,y>$ of $n^{th}$ NN after projection, {$\alpha_i$} represent all combinations of valence and conduction bands, $u_{xk,\alpha}$ is the $x$ component of $\alpha$-band Bloch’s wavefunction calculated at reciprocal point $k$, and $N$ is the finite system size, i.e., the number of allowed reciprocal lattice points. We consider all three valence and conduction bands (Fig. 1(b)) including the two FBs for our calculations.The basis states are $\prod_{cond}\prod_{k'}c_{cond,k'}^{\dagger}\prod_{val}\prod_kc_{val,k}\vert0>$, where $\vert0>$ is defined as completely filled (empty) valence (conduction) bands, and $c_{cond (val),k}^{\dagger}$ creates an electron at reciprocal lattice point $k$ in the conduction (valence) band labelled as cond (val). Eqn.~\ref{eqn2}, therefore, includes all the inter- and intra-band interactions from which we neglect the energetically unfavorable \cite{85} Coulomb induced excitations. Since we are working in the excitonic subspace ($\prod_{cond}\prod_{k'}c_{cond,k'}^{\dagger}\prod_{val}\prod_kc_{val,k}\vert0>$), there are only 6 terms that determine the excitonic interaction. These terms are – 
\begin{enumerate}
    \item Direct e-h interaction: 
$-c_{val,k_3}c_{cond,k_1}^{\dagger}c_{cond,k_4}c_{val,k_2}^{\dagger}$, $-c_{val,k_4}c_{cond,k_2}^{\dagger}c_{cond,k_3}c_{val,k_1}^{\dagger}$
    \item Exchange e-h interaction:
 $c_{val,k_4}c_{cond,k_1}^{\dagger}c_{cond,k_3}c_{val,k_2}^{\dagger}$, $c_{val,k_3}c_{cond,k_2}^{\dagger}c_{cond,k_4}c_{val,k_1}^{\dagger}$
    \item e-e repulsion in conduction band: $c_{cond,k_1}^{\dagger}c_{cond,k_2}^{\dagger}c_{cond,k_3}c_{cond,k_4}$
    \item h-h repulsion in valence band: $c_{val,k_3}c_{val,k_4}c_{val,k_1}^{\dagger}c_{val,k_2}^{\dagger}$
\end{enumerate}
Note that creation operators for electrons in valence band are the destruction operators for holes. The conduction and valence bands pair can be any of the six pair bands in the band structure. The interactions conserves the number of electrons (holes) in conduction (valence) bands ($N_{eh}$) and allows us to solve the Hamiltonian for each $N_{eh}$ separately.\par
Since spin-orbit coupling in our system is negligible, for the calculation of triplet excitons, electron-hole exchange interaction is set to zero, while for singlet excitons, both direct and exchange electron-hole interactions are considered \cite{64}. This follows from the exchange interaction matrix element as given in \cite{64},
\begin{equation}
    x=<vc|K^{ex}|v'c'>=\int dx dx'\psi^*_c(x)\psi_v(x)U(r,r')\psi^*_{c'}\psi_{v'}
\end{equation}
where $v(c)$ labels the valence (conduction) bands, $K^{ex}$ represents exchange interaction in electron-hole kernel \cite{64}. Since Coulomb interaction is spin blind, the spins of v and c must be the same. Thus, we get the exchange interaction matrix as, 
\begin{equation}
    \begin{pmatrix}
    x & x\\
    x & x
    \end{pmatrix}
    \begin{matrix}
    |v\downarrow c\downarrow>\\
    |v\uparrow c\uparrow>
    \end{matrix}
\end{equation}
which clearly shows that the exchange interaction energy for triplet exciton ($|v\uparrow c\uparrow>-|v\downarrow c\downarrow>$) is zero while that of singlet exciton ($|v\uparrow c\uparrow>+|v\downarrow c\downarrow>$) is 2$x$. Hence we have omitted the spin indices in our Hamiltonian for readability. All calculations are performed for a system size $L_x\times L_y=2\times3$. Information about Hilbert space dimensions and convergence can be found in supplementary section II.
\subsection*{Exact Diagonalization (ED) method for solving many-exciton Hamiltonian}
ED method, used for calculating many-body wavefunctions and energies, is known for its computationally expensive nature, both in terms of time and memory \cite{86}. Therefore, it heavily relies on the use of high-performance computational architecture. This method has been previously used for studying mostly fractional Chern insulators \cite{74,76}, where many-body basis states comprise of the possible ways a fixed number of electrons can partially fill the topological FB.\par
In this work we use the same methodology but extend it to more than one band with electrons (holes) in the conduction (valence) bands. Since our projected Hubbard interaction terms conserve the number of electrons (holes) in conduction (valence) bands ($N_{eh}$) as well as the total excitonic momentum, as can be seen from Eqn.~\ref{eqn2}, we use these symmetries to reduce the dimensions of our Hamiltonian. We do our calculations for total excitonic momenta equal to zero block and solve the Hamiltonian for multiple $N_{eh}$. To illustrate our methodology, here we use a fictitious system with one valence and one conduction band, and solve the many-exciton problem with $N_{eh}=2$. We opt a system size of $3\times1$ which implies there are 3 allowed reciprocal lattice momenta: 0, $2\pi/3$, and $4\pi/3$, and work with excitonic population (EP) = $2/(3\times1)$ = 0.67. A typical ED method involves three steps as described below.\par
\textbf{Basis states formation}: The basis states in the many-body Hilbert space for our model are given by $\prod_{k'}c_{c,k'}^{\dagger}\prod_kc_{v,k}\vert0>$, where $\vert0>$ is defined as completely filled (empty) valence (conduction) bands, and $c_{c(v),k}^{\dagger}$ creates an electron at reciprocal lattice point $k$ in the conduction (valence) band labelled as $c(v)$. For our fictitious system, there are $C_2^3$  possible ways that 2 electrons can occupy 3 allowed reciprocal momenta in the conduction band. Similarly, for 2 holes in the valence band there are $C_2^3$ possible combinations. For the block with total excitonic momenta = 0, the basis set comprises of 3 states: 
\begin{align}
&c_{c,2\pi/3}^{\dagger}c_{c,0}^{\dagger}c_{v,2\pi/3}c_{v,0}\vert0>,\nonumber \\
&c_{c,4\pi/3}^{\dagger}c_{c,0}^{\dagger}c_{v,4\pi/3}c_{v,0}\vert0>,\nonumber \\
&c_{c,4\pi/3}^{\dagger}c_{c,2\pi/3}^{\dagger}c_{v,4\pi/3}c_{v,2\pi/3}\vert0>. \nonumber
\end{align}
The ordering of creation and annihilation operators are kept consistent throughout. Our ED code also employs efficient lookup tables for the basis states as their number can reach $\sim10^8$.\par
\textbf{Hamiltonian matrix element computation}: Once the basis set is created, the next step is to construct the many-body Hamiltonian matrix. This requires operating all the terms in the bands-projected Hubbard interaction (Eqn.~\ref{eqn2}) on each basis state. Parallel implementation and memory-mapped I/O are used to store the upper half of this Hermitian matrix on disk, which can reach $\sim1$ TB in disk space. An example of one of the terms acting on one of the basis states for the fictitious system is 
\begin{align}
-(c_{v,2\pi/3}c_{c,0}^{\dagger}c_{c,0}c_{v,2\pi/3}^{\dagger})c_{c,2\pi/3}^{\dagger}c_{c,0}^{\dagger}c_{v,2\pi/3} c_{v,0}\vert0>=-c_{c,2\pi/3}^{\dagger}c_{c,0}^{\dagger}c_{v,2\pi/3}c_{v,0}\vert0>.\nonumber
\end{align}
Here we have used fermionic commutation relation. The interaction term used in this example is the electron-hole direct interaction term which gives a diagonal matrix element.\par
\textbf{Diagonalizing the Hamiltonian}: After the matrix is constructed and stored on disc, we use Lanczos algorithm \cite{59} to find the first few lowest eigenvalues of Hamiltonian. We also compute the wavefunctions and store them for later analysis.

\subsection*{BEC-ED wavefunction overlap}
For case of bosons, since a macroscopic number of particles can occupy the same state, a condensate state would be a properly symmetrized product of single particle state up to the number of particles. In second quantization, this superfluidic state can be written as \cite{55},
\begin{equation}
    |\Phi_{SF}>=(a_i^{\dagger})^N|0>,
\end{equation}
where $a_i^{\dagger}$ creates a boson in the single particle ground state. We can similarly construct such a state for excitons as given in Eqn. 3 using the single triplet excitonic wavefunction. For the fictitious system, we assume that the single triplet ground state (calculated with $N_{eh}=1$) excitonic wavefunction is given by 
\begin{align}
b_{exc}^{\dagger}=\frac{1}{\sqrt{3}}(c_{c,0}^{\dagger}c_{v,0}+c_{c,2\pi/3}^{\dagger}c_{v,2\pi/3}+c_{c,4\pi/3}^{\dagger}c_{v,4\pi/3}).\nonumber
\end{align}
Using this wavefunction, an ideal BEC two-particle wavefunction of the form given by Eqn. 3 can be constructed as
\begin{align}
\vert\phi_{BEC}>&\sim[b_{exc}^{\dagger}]^2\vert0>\sim(c_{c,0}^{\dagger}c_{v,0}+c_{c,2\pi/3}^{\dagger}c_{v,2\pi/3}+c_{c,4\pi/3}^{\dagger}c_{v,4\pi/3})^2\vert0>\nonumber\\
&=(c_{c,0}^{\dagger}c_{v,0}+c_{c,2\pi/3}^{\dagger}c_{v,2\pi/3}+c_{c,4\pi/3}^{\dagger}c_{v,4\pi/3})\nonumber\\
&\times(c_{c,0}^{\dagger}c_{v,0}+c_{c,2\pi/3}^{\dagger}c_{v,2\pi/3}+c_{c,4\pi/3}^{\dagger}c_{v,4\pi/3})\vert0>.\nonumber
\end{align}
Since for fermions $c^{\dagger}c^{\dagger}\vert\psi>=0$,
\begin{align}
\vert\phi_{BEC}>&\sim(c_{c,0}^{\dagger}c_{v,0}c_{c,2\pi/3}^{\dagger}c_{v,2\pi/3}+c_{c,0}^{\dagger}c_{v,0}c_{c,4\pi/3}^{\dagger}c_{v,4\pi/3}\nonumber\\
&+c_{c,2\pi/3}^{\dagger}c_{v,2\pi/3}c_{c,0}^{\dagger}c_{v,0}+c_{c,2\pi/3}^{\dagger}c_{v,2\pi/3}c_{c,4\pi/3}^{\dagger}c_{v,4\pi/3}\nonumber\\
&+c_{c,4\pi/3}^{\dagger}c_{v,4\pi/3}c_{c,0}^{\dagger}c_{v,0}+c_{c,4\pi/3}^{\dagger}c_{v,4\pi/3}c_{c,2\pi/3}^{\dagger}c_{v,2\pi/3})\vert0>.\nonumber
\end{align}
Next, using fermionic commutation relations,
\begin{align}
\vert\phi_{BEC}>\sim2(c_{c,0}^{\dagger}c_{v,0}c_{c,2\pi/3}^{\dagger}&c_{v,2\pi/3}+c_{c,0}^{\dagger}c_{v,0}c_{c,4\pi/3}^{\dagger}c_{v,4\pi/3}\nonumber\\
&+c_{c,2\pi/3}^{\dagger}c_{v,2\pi/3}c_{c,4\pi/3}^{\dagger}c_{v,4\pi/3})\vert0>.\nonumber
\end{align}
We need to reorder the creation and annihilation operators in accordance with the chosen ordering of basis states as mentioned above,
\begin{align}
\vert\phi_{BEC}>\sim2(-c_{c,2\pi/3}^{\dagger}c_{c,0}^{\dagger}&c_{v,2\pi/3}c_{v,0}-c_{c,4\pi/3}^{\dagger}c_{c,0}^{\dagger}c_{v,4\pi/3}c_{v,0}\nonumber\\
&-c_{c,4\pi/3}^{\dagger}c_{c,2\pi/3}^{\dagger}c_{v,4\pi/3}c_{v,2\pi/3})\vert0>.\nonumber
\end{align}
In order to calculate the BEC-ED overlap, we use this ideal BEC two-particle wavefunction and the ED ground-state triplet wavefunction calculated with $N_{eh}=2$.

\subsection*{Two-body density matrix}
The eigenvalues of the reduced two-body density matrix can be used to show the degree of condensation in many-body excitonic wavefunction. For interacting bosonic systems, the condensation criterion was established by Penrose and Onsager \cite{66} with the existence of one very large eigenvalue of the density matrix. This is particularly useful when the one-body ground state, into which the bosons condense, cannot be known \textit{a priori}. The eigenvalues of density matrix are the occupations of ‘true’ orbitals of the system, given by its eigenfunctions. Hence, the existence of one very large eigenvalue unambiguously illustrates condensation and coherence. Criterion for condensation of composite fermions (Cooper pairs of superconductivity) was given by Yang \cite{67} as an extension to the Penrose and Onsager criterion \cite{66}. It was shown that the existence of one large eigenvalue of the reduced two-body density matrix is related to the emergence of off-diagonal long-range order, a fundamental and central characterization of superfluidity. Hence, for a system of $N$ fermions there should exist one eigenvalue of ‘order $N$’, while others of order unity if the cooper pairs condense, while if there are more than one eigenvalues of ‘order $N$’, the system becomes fragmented \cite{68}.\par
We can similarly formulate a reduced two-body density matrix for electron-hole systems as given by Eqn. 2. There is a subtle difference between the density matrix for electron-electron systems \cite{67} and the one we formulate here. For the case of Cooper pairs, both electrons have access to the same single-particle states due to which the summation of eigenvalues of the reduced two-body density matrix is constrained to $N(N-1)$ where $N$ is the number of fermions in the system. In contrast, the summation of eigenvalues for electron-hole system considered here should be $N_{eh}^2$ since the electrons (holes) have access to the states in conduction (valence) bands independently. The condition of Yang criterion \cite{67} remains the same since the formation of superfluid of correlated electrons and holes should be characterized by the existence of ODLRO.\par
In our methodology, we define the density matrix as,
\begin{align}
\rho^{(2)}(k,k';\overline{k},\overline{k}')=<\Psi\vert c_{c,k}^{\dagger}c_{v,k'}c_{v,\overline{k}'}^{\dagger}c_{c,\overline{k}}\vert\Psi>,\nonumber
\end{align}
where $\vert\Psi>$ is the excitonic many-body wavefunction. Each of the $k$, $k'$, $\overline{k}$, and, $\overline{k}'$ can be any of the allowed reciprocal lattice vectors. Fermionic commutation relations lead to certain matrix elements being related to each other, which reduces the dimension of matrix in one direction to be equal to the square of the number of allowed $k$-points.\par
For our fictitious system setup, we have 3 allowed $k$-points, 1 conduction band, and 1 valence band. This implies that the dimensions of density matrix would be $9\times9$. As an example, we illustrate here the calculation of one of the matrix elements of density matrix. Assume a many-body wavefunction given by $\vert\Psi>=c_{c,2\pi/3}^{\dagger}c_{c,0}^{\dagger}c_{v,2\pi/3}c_{v,0}\vert0>$,
\begin{align}
<\Psi\vert c_{c,4\pi/3}^{\dagger}&c_{v,2\pi/3}c_{v,2\pi/3}^{\dagger}c_{c,0}\vert\Psi>\nonumber\\
&=<\Psi\vert c_{c,4\pi/3}^{\dagger}c_{v,2\pi/3}c_{v,2\pi/3}^{\dagger}c_{c,0}c_{c,2\pi/3}^{\dagger}c_{c,0}^{\dagger}c_{v,2\pi/3}c_{v,0}\vert0>\nonumber\\
&=-<\Psi\vert c_{c,4\pi/3}^{\dagger}c_{v,2\pi/3}c_{v,2\pi/3}^{\dagger}c_{c,2\pi/3}^{\dagger}c_{v,2\pi/3}c_{v,0}\vert0>\nonumber\\
&=-<\Psi\vert c_{c,4\pi/3}^{\dagger}c_{c,2\pi/3}^{\dagger}c_{v,2\pi/3}c_{v,0}\vert0>\nonumber\\
&=-<0\vert c_{v,0}^{\dagger}c_{v,2\pi/3}^{\dagger}c_{c,0}c_{c,2\pi/3}c_{c,4\pi/3}^{\dagger}c_{c,2\pi/3}^{\dagger}c_{v,2\pi/3}c_{v,0}\vert0>=0.\nonumber
\end{align}
This matrix element is zero since $c_{c,0}\vert0>=0$.
\section{Hilbert space dimensions and convergence with respect to finite system size}
In this section we show the convergence of our results with respect to finite system size. Since we calculate different EPs, the Hilbert space size varies with EP reaching maxima at the completely population inverted (CPI) state. This CPI state is defined as one with $N_{eh}$ equal to the total number of allowed reciprocal lattice points. In case of a $2\times3$ finite suze lattice, as used, in this work, CPI state is the one with $N_{eh}=6$. To be consistent and accurate, one needs to use the same finite system size for all excitonic populations. Hence, we are bottle-necked by the CPI Hilbert space dimensions for a given finite system size. In Table~\ref{tab:tables1}, we show the maximum Hilbert space size for various finite system sizes. For our calculations we use a 2×3 system size which includes 36 lattice sites. In order to solve the CPI state for this finite system, we used a paralleled Julia code written in-house that ran over 180 nodes each with 15 threads (total cpus = 2700) and took ~240 hours to do all the analysis for one system.\par
\begin{figure}[H]
\includegraphics[scale=0.5]{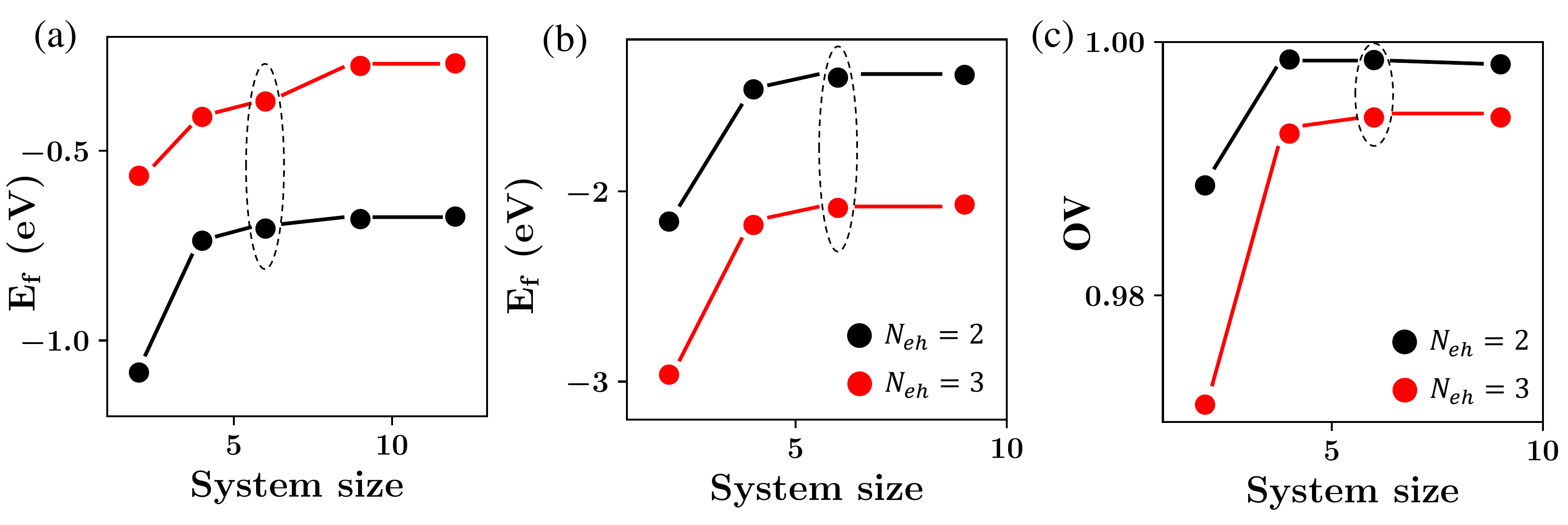}
\caption{\label{fig:figs1} \textbf{Convergence test results with respect to finite system size.} (a) Formation energies of the first two single triplet excitonic levels calculated for $EI_{FB}$; black circle denotes the first excitonic level while red circle denotes the second, (b) Formation energies of excitonic levels with $N_{eh}>1$ calculated for $EI_{FB}$, and (c) BEC-ED wavefunction overlaps for $N_{eh}>1$  calculated for $EI_{FB}$. In all cases, our results are well converged for a system size of $2\times3$, as marked by dotted ovals.}
\end{figure}

In the following, we illustrate our convergence results. In Fig.~\ref{fig:figs1}(a), we show the convergence of single exciton formation energies of the first two triplet exciton levels of $EI_{FB}$ with system size. Clearly, our chosen system size, 2×3 is converged and allows us to be consistent in our choice of finite system size for multiple EPs. In Fig.~\ref{fig:figs1}(b), we show convergence of formation energies of first tow triplet exciton levels of $EI_{FB}$ for $N_{eh}>1$. Finally, we check for convergence of many-body wavefunction properties, which is shown in Fig.\ref{fig:figs1}(c). From the figures one can see that our calculations are converged, and our results are well within the acceptable numerical error.\par
\section{Benchmark ED results with GW-BSE}
\begin{figure}[H]
\includegraphics[scale=0.5]{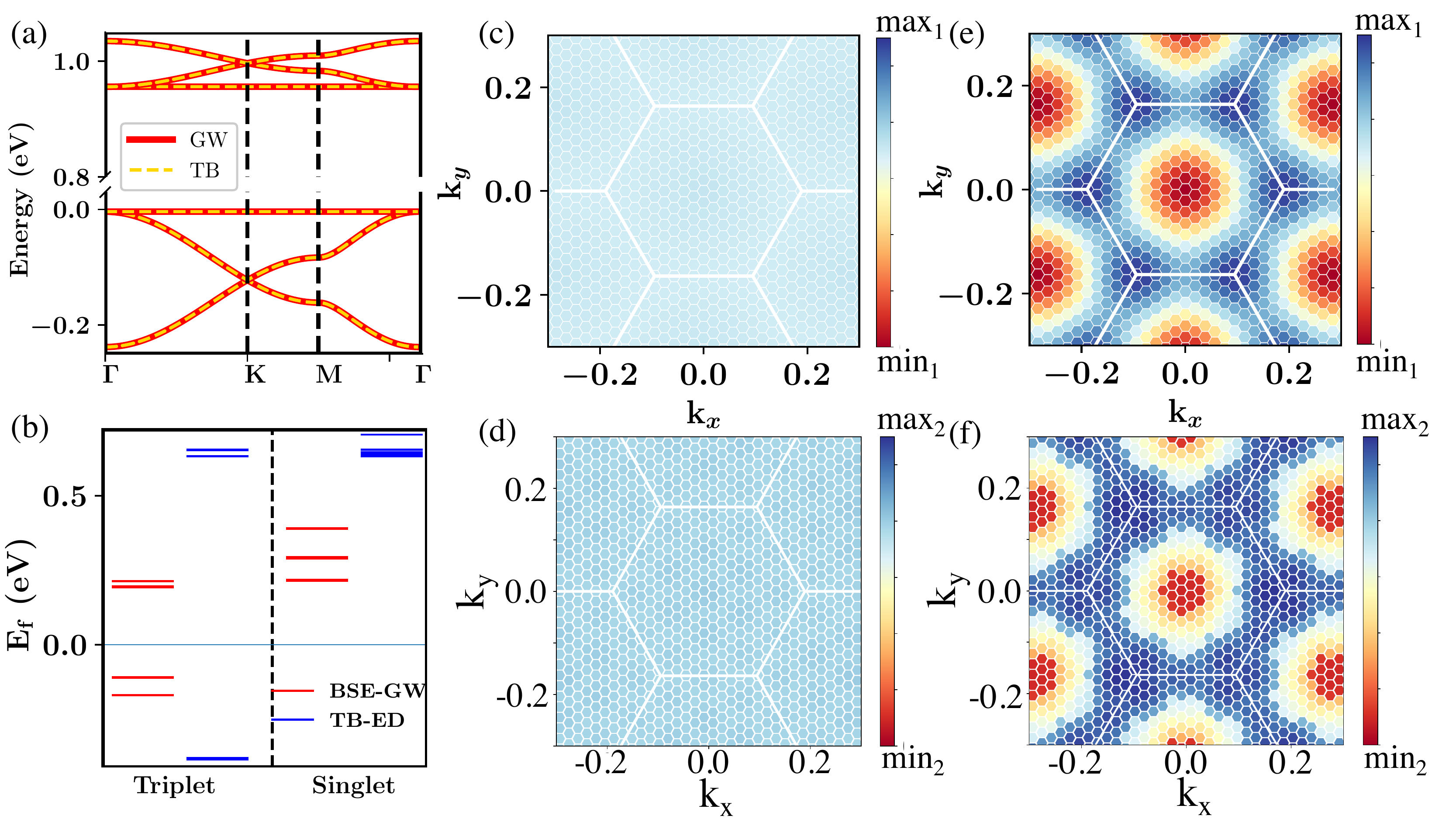}
\caption{\label{fig:figs2} \textbf{TB-ED benchmarked with BSE-GW.} (a) Tight binding fit of the GW band structure calculated for $EI_{SG}$, (b) Exciton formation energy of TB-ED results (blue bars) compared with GW-BSE calculations (red bars) for $EI_{SG}$. (c) First triplet excitonic wavefunction of $EI_{SG}$ calculated using ED, (d) First triplet excitonic wavefunction of $EI_{SG}$ obtained using GW-BSE, (e) First singlet excitonic wavefunction of $EI_{SG}$ calculated using ED, (f) First singlet excitonic wavefunction of $EI_{SG}$ obtained using GW-BSE. The ED calculated wavefunctions exactly match with the ones obtained using BSE.}
\end{figure}

In Fig.~\ref{fig:figs2}(a), we show the tight-binding fit of GW band structure. A perfect fit was obtained with tight-binding hopping parameters, $t_1=0.532$, $t_2=0.0258$, and $t_3=0.0261$. Using these parameters in the ED method, we benchmarked the exciton formation energies against the ones we got from GW-BSE as shown in Fig. 1(b) in the main text, which is reproduced in Fig. ~\ref{fig:figs2}(b). We also benchmarked excitonic wavefunctions for the $9\times9$ superatomic graphene lattice ($EI_{SG}$) \cite{6}. In Fig.~\ref{fig:figs2}(c) [Fig.~\ref{fig:figs2}(e)] and Fig.~\ref{fig:figs2}(d) [Fig.~\ref{fig:figs2}(f)], we show a perfect match of the first triplet [singlet] wavefunctions, calculated using ED and BSE, respectively.\par
\begin{figure}[H]%
\centering
\includegraphics[width=0.8\textwidth]{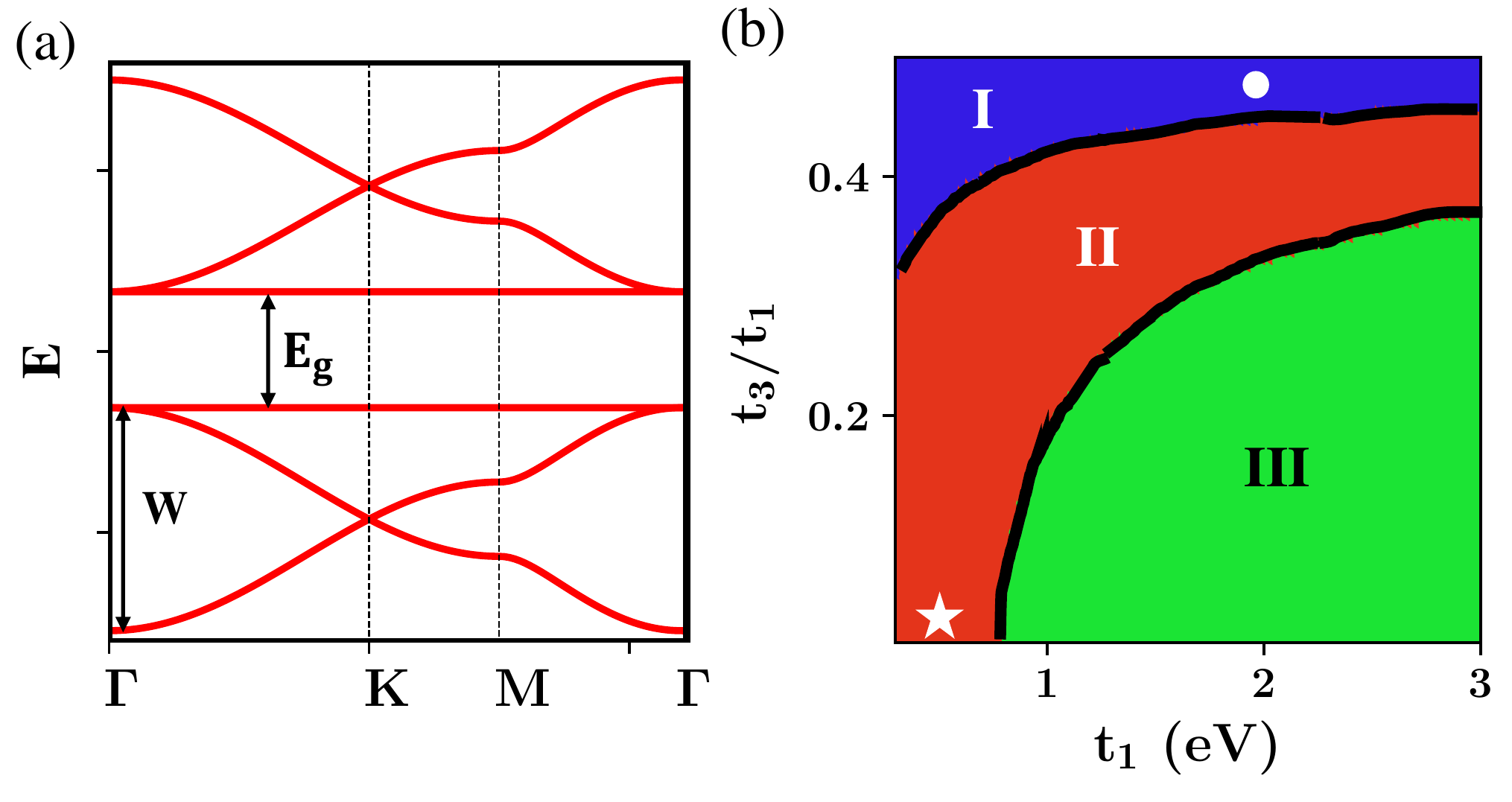}
\caption{\label{fig:figs3} \textbf{Phase diagram of single triplet exciton stability:} (a) TB band structure of diatomic Kagome lattice with yin-yang FBs. The dependence of $E_g$ and band width (W) on hopping parameters ($t_2$ is set to zero for simplicity) are $E_g = 2t_1-4t_3$ and $W = 6t_3$. (b) Phase diagram showing three distinct regimes of different exciton stability. Star marks the system $EI_{SG}$ while dot represents $EI_{FB}$.}
\end{figure}

We also mapped a phase diagram in the inter-atomic hopping integrals set when $t_2$ is set to zero as depicted in Fig. ~\ref{fig:figs3}(b). There are three regimes in the phase diagram: in regime I (blue), the $E_b$ of both singlet and triplet excitons exceeds the $E_g$; in regime II (red), only triplet $E_b$ exceeds $E_g$; in regime III (green), neither singlet nor triplet $E_b$ exceeds $E_g$. In Fig. ~\ref{fig:figs3}(a), where we plot an ideal yin-yang band structure \cite{59}, we indicate the dependence of $E_g$ and band width (W) on hopping integrals which helps explain the three regimes. For example, for a fixed $t_1$, as $t_3$ increases, $E_g$ decreases while W increases. This leads to increased contribution of FBs to the excitonic levels which is the case for $EI_{FB}$ denoted on the phase diagram with dot. We have also denoted $EI_{SG}$ with a star in the phase diagram. Note that $EI_{PB}$ doesn't lie in this phase diagram since it has non-zero $t_2$.
\section{Eigenvalue spectra of the two-body density matrix}
\begin{figure}[H]
\centering
\includegraphics[width=0.9\textwidth]{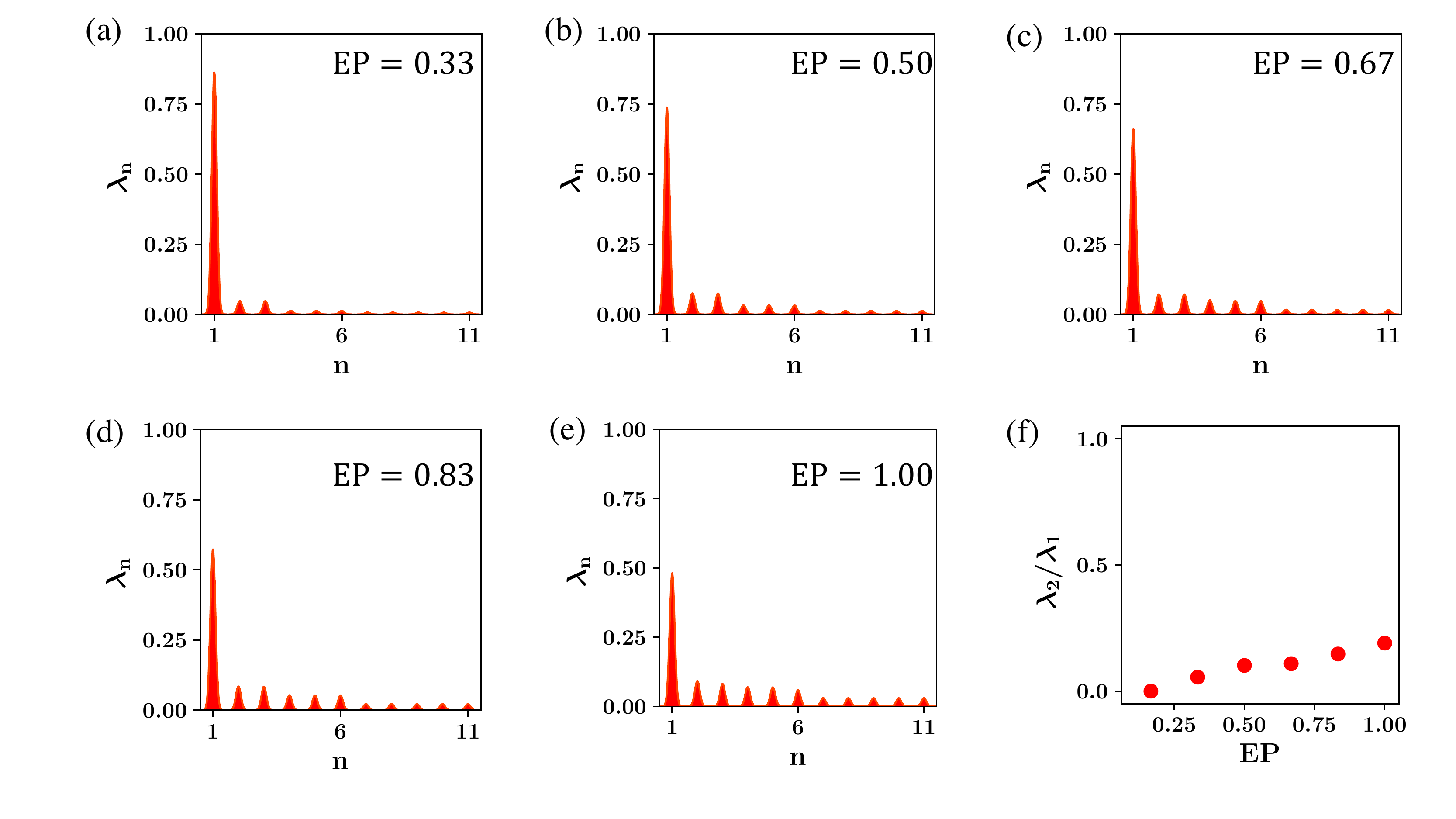}
\caption{\label{fig:figs5} \textbf{Eigenvalue spectra of the two-body density matrix for $EI_{SG}$}: (a) – (e) First few largest normalized eigenvalues ($\lambda_n$) of reduced two-body density matrix calculated for ground state many-triplet-excitonic wavefunctions at various EPs. A BEC condensate is formed for all EPs but in a different form at low vs. high EPs as indicated by the change in the slope of $\lambda_2/\lambda_1$ shown in (f).}
\end{figure}

\begin{figure}[H]
\centering
\includegraphics[width=0.9\textwidth]{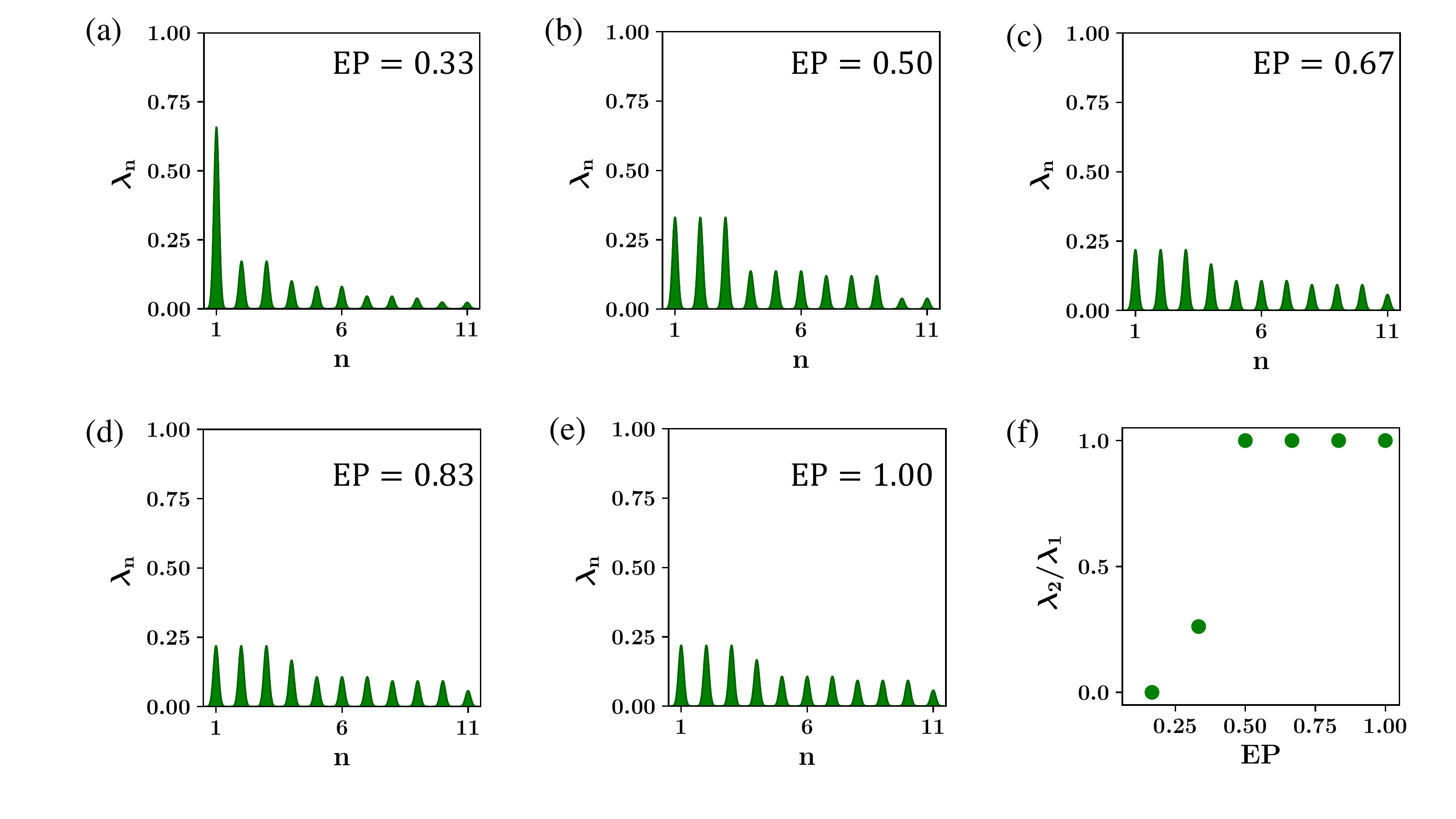}
\caption{\label{fig:figs6} \textbf{Eigenvalue spectra of the two-body density matrix for $EI_{PB}$}: (a) – (e) First few largest normalized eigenvalues ($\lambda_n$) of reduced two-body density matrix calculated for ground state many-triplet-excitonic wavefunctions at various EPs. Triplet excitons in this case are not condensing. The ratio $\lambda_2/\lambda_1$ plotted in (f) shows complete fragmentation for EP$>$0.33}
\end{figure}

\begin{figure}[H]
\centering
\includegraphics[width=0.9\textwidth]{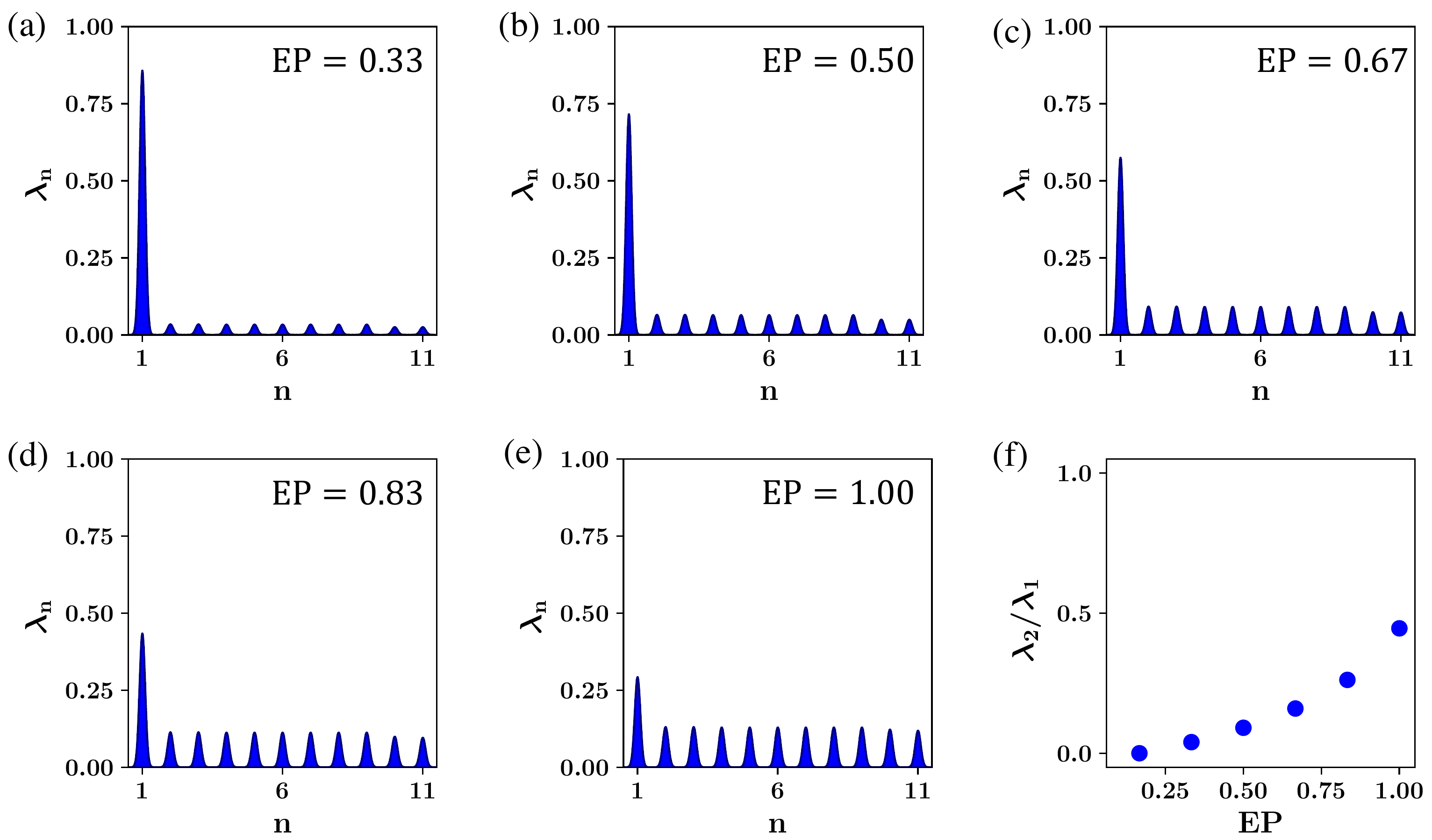}
\caption{\label{fig:figs4} \textbf{Eigenvalue spectra of the two-body density matrix for $EI_{FB}$:} (a)-(e) First few largest normalized eigenvalues ($\lambda_n$) of reduced two-body density matrix calculated for ground state many-triplet-excitonic wavefunctions at various EPs. There exists one leading eigenvalue close to 1 in all cases, signifying the BEC for all EPs up to 0.83 shown in (d). In (f) the ratio $\lambda_2/\lambda_1$ is plotted. There is some degree of condensation even for the CPI state, as indicated by $\lambda_2/\lambda_1\sim0.5$ at EP$=1$.}
\end{figure}

We plot the complete eigenvalue spectra of the two-body density matrix for ground-state wavefunctions of $EI_{SG}$, $EI_{PB}$, and $EI_{FB}$ in Fig.~\ref{fig:figs5}, Fig.~\ref{fig:figs6}, and Fig.~\ref{fig:figs4} respectively.\par
In the case of $EI_{SG}$ (Fig.~\ref{fig:figs5}) for lower EPs, the decreasing trend of $\lambda_2/\lambda_1$ matches well with the overlap trend (red crosses in Fig. 3(b)); while for EP above 0.5, there exists one large eigenvalue even though the BEC-ED overlap in Fig. 3(b) is quite low. This implies that for higher EPs excitons are still forming a condensate, but in a form different than Eqn. (3), which is also indicated by the change in the slope of $\lambda_2/\lambda_1$ as shown in Fig. Fig.~\ref{fig:figs5}(f).\par
For the case of $EI_{PB}$, the ground state many-excitonic wavefunctions are not a condensate at any EP as shown in Fig.~\ref{fig:figs6}.\par
\begin{figure}[H]
\centering
\includegraphics[width=0.5\textwidth]{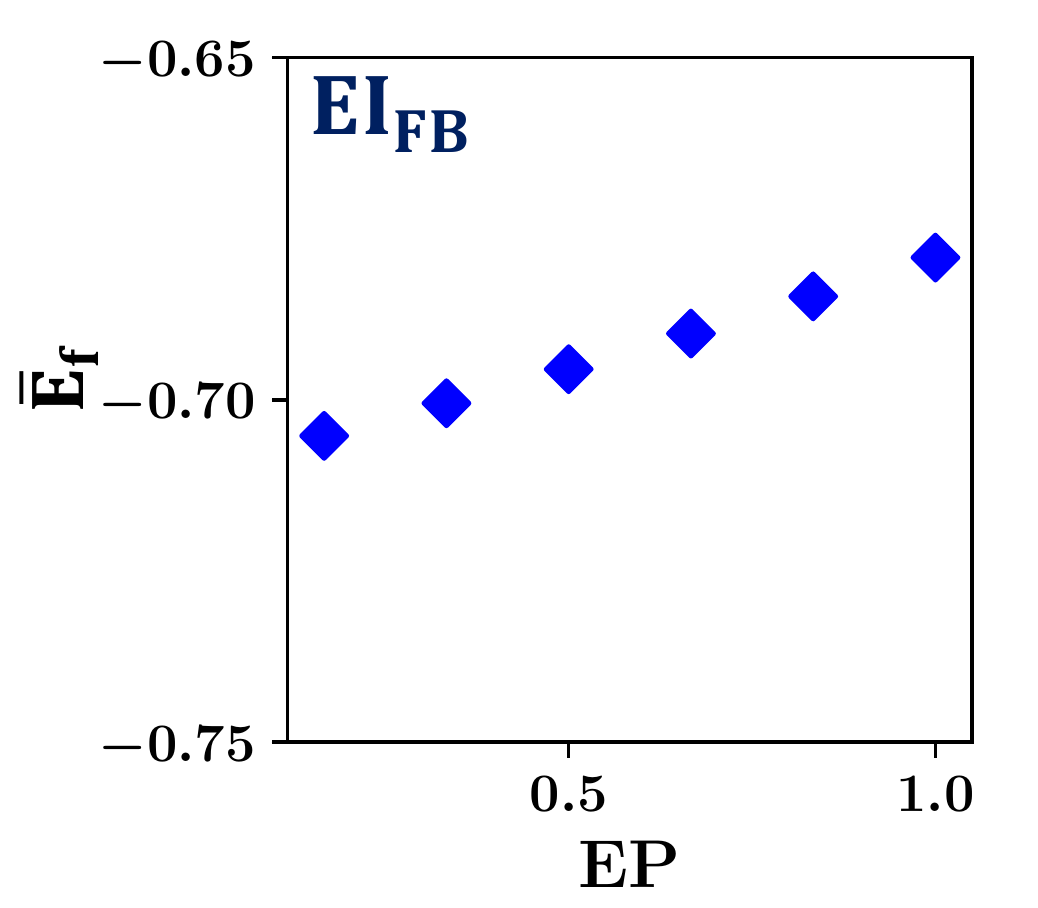}
\caption{\label{fig:figs7} \textbf{Many-excitonic ground state energies at multiple EPs for $EI_{FBB}$}: This plot has the same scale as that of Fig. 2(a) and 2(b). It can be seen that the ground state triplet excitons of $EI_{FB}$ experience a slightly larger exciton-exciton repulsion than $EI_{SG}$ ;$\bar{E}_f$ increases by 3.6$\%$ from EP = 0.17 to EP = 1.0. The interactions are still very weak suggesting that the triplet excitons in this case might be condensing as well. }
\end{figure}

We also plot the complete $\lambda_n$ spectra of the density matrix for the ground-state triplet wavefunctions of $EI_{FB}$ at multiple EPs in Fig. Fig.~\ref{fig:figs4}. Clearly, one sees $\lambda_1$ of the order 1 existing for EPs up to 0.67, which is a signature of triplet EBEC. As EP increases, lower eigenvalues begin to rise a bit indicating a slight fragmentation of the condensate. In Fig. ~\ref{fig:figs4}(f), we plot the $\lambda_2/\lambda_1$ ratio for ground-state wavefunctions. One sees that there is still some degree of condensation in the CPI (EP = 1) state with $\lambda_2/\lambda_1 ~ 0.5$. This agrees with the trend of BEC-ED overlaps as shown in Fig. 3(b) indicating that triplet excitons in this case are condensing to a BEC state as given in Eqn. (3). This is also consistent with the multiple exciton formation energies for $EI_{FB}$ as plotted in Fig. ~\ref{fig:figs7}\par


\section{Supplementary Tables}
\begin{table}[H]
\caption{\label{tab:tables1}%
Hilbert space dimensions for various finite system sizes. In this work we work with a system of size $2\times3$.
}
\begin{ruledtabular}
\begin{tabular}{cc}
\textrm{System size} & \textrm{CPI Hilbert space dimensions}\\
\colrule
$2\times2$ & $245,025$ \\
$2\times3$ & $344,622,09$ \\
$2\times4$ & $540,917,591,841$ \\
$3\times3$ & $21.99\times10^{12}$\\
\end{tabular}
\end{ruledtabular}
\end{table}


\section{Supplementary Figures}

\bibliography{manuscript}

\end{document}